# Simplified equivalent circuit approach for designing time-domain responses of waveform-selective metasurfaces

Kosei Asano,[1,a)] Tomoyuki Nakasha,[1,a)] and Hiroki Wakatsuchi[1,2,b)]

[1]*Department of Electrical and Mechanical Engineering, Graduate School of Engineering, Nagoya Institute of*

*Technology, Aichi, 466-8555, Japan*

[2]*Precursory Research for Embryonic Science and Technology (PRESTO), Japan Science and Technology Agency (JST),*

*Saitama 332-0012, Japan*

In this study, we demonstrate simplified equivalent circuit models to effectively approximate responses of recently reported waveform-selective metasurfaces that distinguish different electromagnetic waves even at the same frequency depending on their waveforms or pulse widths. Compared to conventional equivalent circuit models that represent behaviors of ordinary metasurfaces in the "frequency" domain, the proposed models enable us to explain how waveform-selective metasurfaces respond in the "time" domain. Our approach well estimates not only time constants of waveform-selective metasurfaces but also their entire time-domain responses. Particularly, this study reports the importance of resistive components of diodes as well as a limitation of our models with respect to power dependence, although still the models effectively work when waveform-selective mechanisms are clearly exhibited with a sufficiently large input power. Thus, the idea of the proposed equivalent circuit models contributes not only to more understanding waveform-selective mechanisms but also to facilitating the design process of such unique structures.

**THE MANUSCRIPT**

In electromagnetics, artificially engineered periodic structures[1–5] are well known to have an advantage over conventional materials available in nature, since their electromagnetic responses are readily tailored by adjusting their subwavelength geometry to produce various electromagnetic characteristics including negative/zero refractive index[6,7] and extremely large surface impedance[3]. Particularly, a planar type of structures, or the so-called metasurfaces, has a simpler form than 3D periodic structures and is therefore used for a wide range of applications such as wavefront shaping[8–10], spatial filtering screens[11–13], RF (radio-frequency) or optical absorbers[14–16], and digital coding[17]. Moreover, these unique properties and performances are further extended by introducing a nonlinearity to a metasurface[18–22]. Although ordinary metasurfaces have strong frequency dependence and are thus designed to operate at the frequency of interest, circuit-based metasurfaces containing schottky diodes were recently demonstrated to be capable of distinguishing different waves in response not only to the incoming frequency components but also to the waveforms or pulse widths (FIG. 1(a))[23–27]. Hence, these waveform-selective metasurfaces were expected to give us an additional degree of freedom to control electromagnetic waves even at the same frequency. For instance, the use of these materials enabled us to effectively absorb an arbitrary waveform and lower its bit error rate compared to other signals at the same frequency, by deploying a proper type of circuit configuration[24,28]. However, although these unique characteristics are easily changed by circuit constants used, none of past studies has yet to clarify how waveform-selective mechanisms are related to the circuit components. For over a decade, equivalent circuit models have been used as a useful approach to understand how responses of conventional metasurfaces were determined by fine features of periodic unit cells, which were effectively approximated by capacitors, inductors, resistors, and so on[29–31]. While this approach facilitates the design process of metasurfaces, it is applicable to estimating only the frequency-domain response but not waveform-selective response varying in the time domain. Other approaches include, for instance, the interference theory[32,33] and retrieval methods[34,35], both of which, however, characterize electromagnetic behavior or properties at steady state as well. Besides, these methods cannot associate transient response with lumped circuit elements as functions of their circuit constants. For this reason, this study presents an approach based on equivalent circuit models that are applicable to predicting time-domain responses of waveform-selective metasurfaces unlike classic equivalent circuit approaches. Our models are proposed using two simplified assumptions. Nonetheless, they still effectively represent how waveform-selective metasurfaces behave in the time domain including their time constants.

Waveform-selective absorbing mechanisms have been already reported in our previous studies[23–25] but can be briefly understood as below. Firstly, our structures are designed to have a 18 mm periodicity and consist of square conducting patches (perfect electric conductor: PEC; 17×17 mm$^2$), dielectric substrate (1.5-mm-thick Rogers3003 but without loss for the sake of simplicity), and ground plane (FIG. 1(a)). Additionally, each gap between conducting patches is connected by a set of four diodes (Broadcom, HSMS286x series), which, as a diode bridge, fully rectifies induced electric charges and generates an infinite set of frequency components. As expected from the Fourier series expansion of the fully rectified waveform, however, most energy is at zero frequency. Hence, by connecting an inductor and a capacitor to a resistor inside the diode bridge as an inductor-based waveform-selective metasurface (FIG. 1(b)) and a capacitor-based waveform-selective metasurface (FIG. 1(c)), respectively, the electric charges can be temporarily controlled in the time domain even at the same frequency as gradually the inductor lowers its electromotive force and the capacitor is charged up. These time-varying voltages across the inductor and the capacitor are related to absorptance as seen in FIG. 1(d) and FIG. 1(e), which show that effectively the structures have time-varying absorptance unlike conventional metasurfaces[36]. In the following part of this study, the frequency of an incident wave is set to 4.1 GHz where waveform-selective performances of our structures are optimized.

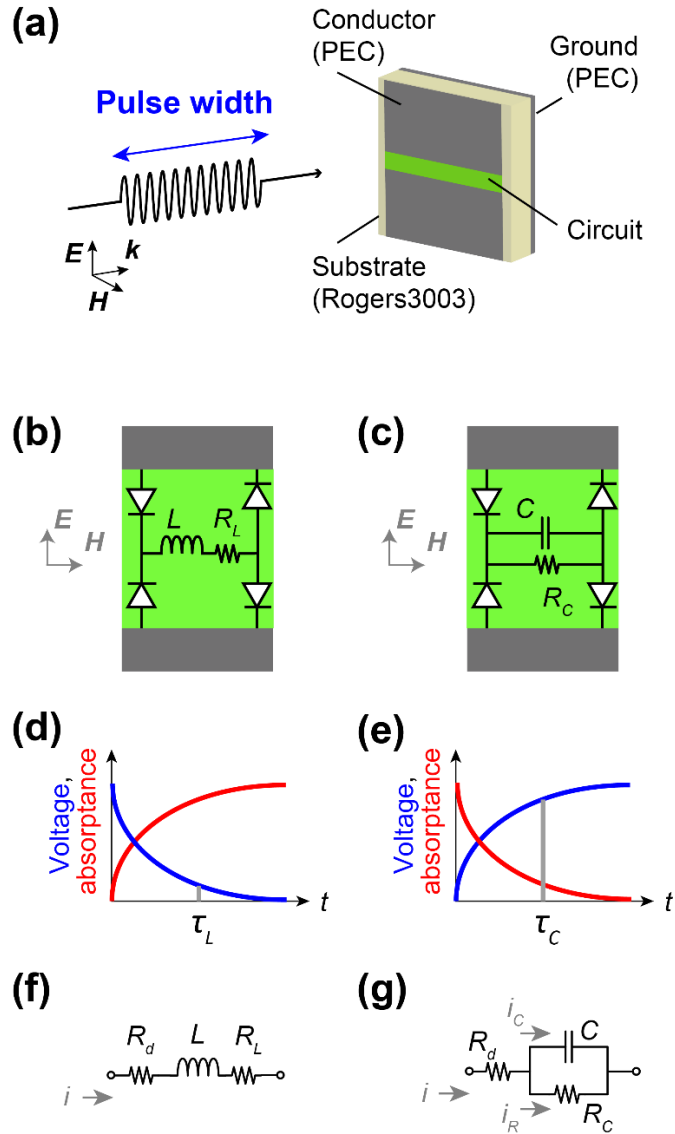

FIG. 1. (a) Pulse width and a periodic unit cell of a waveform-selective metasurface. (b,c) The circuit configurations deployed between patches for an inductor-based waveform-selective metasurface and a capacitor-based waveform-selective metasurface. (d,e) Their time-domain responses and (f,g) proposed equivalent circuit models that simplify the circuit response between patches.

Moreover, combining the two circuit configurations leads to designing more complicated waveform-selective absorbing mechanisms[24]. For example, when the two circuit configurations are connected to each other in parallel as a parallel-type waveform-selective metasurface, both of a short pulse and a continuous wave (CW) are strongly absorbed due to the presence of the capacitor- and inductor-based circuits. However, still the electromotive force of the inductor is not sufficiently lowered by an intermediate pulse, while it charges up the capacitor completely. Therefore, compared to a short pulse and a CW, the parallel-type waveform-selective metasurface poorly absorbs an intermediate pulse, which appears as a voltage peak. In contrast, if these two circuits are combined in series as a series-type waveform-selective metasurface, incoming electric charges induced by a short pulse are blocked by the electromotive force of the inductor, while a CW fully charges up the capacitor. For this reason, this structure more strongly absorbs an intermediate pulse than both a short pulse and a CW, which is shown as a voltage dip.

To approximate all these mechanisms, we firstly developed an equivalent circuit model for an inductor-based waveform-selective metasurface. Since it is very difficult to rigorously incorporate the nonlinear response of schottky diodes, we applied the following two assumptions. As mentioned above, most of the energy of induced electric charges is converted to zero frequency component within a diode bridge. Therefore, we assumed that the circuits deployed in the gap between conducting patches were biased by a DC (direct current) voltage source. Next, schottky diodes vary the degree of the current to come in, which depends on the voltage applied. This means that their effective resistance $R_0$ varies in response to the voltage applied (i.e., $R_0$ changes in the time domain). For the sake of simplicity, however, these diodes were assumed to have a constant resistance that was estimated by the relationship between the turn-on voltage $V_{on}$ and its current $I_{on}$, namely, by $R_0 = V_{on}/I_{on}$. In this study, we used commercial diodes provided by Broadcom (specifically, HSMS286x series) but excluded parasitic circuit parameters such as a series resistance. As a result, $R_0$ was calculated to be 340 Ω.

Under these circumstances, the entire circuit between conducting patches may be effectively represented as drawn in FIG. 1(f). Note that $R_d$ is double of the effective resistance of a single diode (i.e., $R_d = 2R_0$), since rectified electric charges enter two diodes to reach a neighboring patch. If a DC voltage $E_0$ is applied to this circuit, then the following equation is obtained:

$$(R_L + R_d)i(t) + Ldi(t)/dt = E_0, \qquad (1)$$

where $R_L$, $L$, $t$, and $i$ are series resistance, inductance, time, and the current flowing into the entire circuit, respectively. Supposing initial condition $i(0) = 0$, solving this equation with respect to $i$ yields

$$i(t) = E_0\left(1 - e^{-\frac{R_L+R_d}{L}t}\right)(R_L + R_d)^{-1}. \qquad (2)$$

By differentiating eq. (2) with respect to $t$ and multiplying it by $L$, inductor voltage $v_L$ is given by

$$v_L(t) = E_0 e^{-\frac{R_L+R_d}{L}t} = E_0 e^{-\frac{t}{\tau_L}}, \qquad (3)$$

where time constant $\tau_L$ is

$$\tau_L = L/(R_L + R_d). \qquad (4)$$

Similarly, a simplified equivalent circuit model of a capacitor-based waveform-selective metasurface may be represented as shown in FIG. 1(g), and a set of the following circuit equations is derived for this circuit model:

$$\begin{cases} i_C(t) = dq(t)/dt \\ R_C i_R(t) = C^{-1} \int i_C(t) dt \\ R_d\{i_R(t) + i_C(t)\} + R_C i_R(t) = E_0, \end{cases} \quad (5)$$

where $R_C$, $C$, $q$, $i_C$, and $i_R$ denote parallel resistance, capacitance, the charges stored in the capacitor, the current at the capacitor, and that at the resistor, respectively. Using these equations, the capacitor voltage $v_c$ and time constant $\tau_C$ are, respectively, given by

$$v_C(t) = R_C(R_C + R_d)^{-1} E_0 \left(1 - e^{-\frac{R_C + R_d}{CR_C R_d}t}\right) = R_C(R_C + R_d)^{-1} E_0 \left(1 - e^{-\frac{t}{\tau_C}}\right), \quad (6)$$

$$\tau_C = C R_C R_d / (R_C + R_d). \quad (7)$$

If $R_C \gg R_d$ then $\tau_C$ becomes a simpler form of

$$\tau_C \sim C R_d. \quad (8)$$

Note that from eqs. (4), (7), and (8), the time constants of the inductor- and capacitor-based waveform-selective metasurfaces are expected to be proportional to $L$ and $C$, respectively.

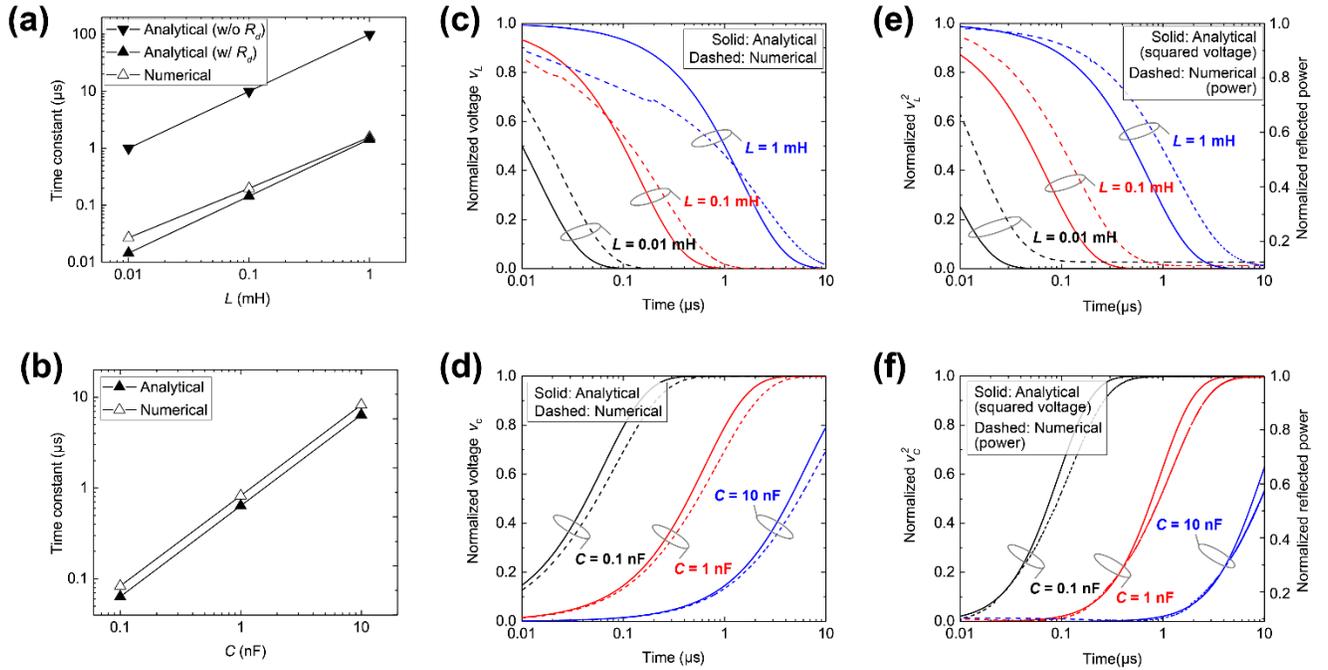

FIG. 2. Comparison in time constants of (a) an inductor-based waveform-selective metasurface and (b) a capacitor-based waveform-selective metasurface. (c,d) Their inductor and capacitor voltages in the time domain. (e,f) Comparison with their reflected powers.

Equations (4) and (7) were used to estimate time constants of these waveform-selective metasurfaces. These analytically derived results were then compared to numerical simulation results that were obtained by a co-simulation method integrating a commercial electromagnetic simulator with a circuit simulator (ANSYS, Electronics Desktop R18.1.0)[23–25]. Firstly, a comparison result for an inductor-based waveform-selective metasurface is shown in FIG. 2(a), where inductance $L$ was swept in a wide range between 0.01 and 1 mH, while resistance $R_L$ was fixed at 10 Ω. In this case, the analytically derived time constant was found to be close to the numerical result by using diode resistance $R_0$ (i.e., $R_d$). This indicates that the diode resistance plays an important role to model the time-domain response of an inductor-based waveform-selective metasurface. Similarly, time constants of a capacitor-based model were compared in FIG. 2(b), where capacitance $C$ was varied from 0.1 to 10 nF, while resistance $R_C$ was set to 10 kΩ. As a consequence, the analytical result closely agreed with the numerical result for the capacitor-based waveform-selective metasurface as well. Moreover, close agreement was obtained, when $R_L = 1$ Ω and 100 Ω and $R_C = 1$ kΩ and 100 kΩ, which cover a practical range of the resistors to achieve good waveform-selective performances and to obtain clear gap between absorptance for a short pulse and that for a CW (results not shown here). This indicates that eqs. (4) and (7) can be used as good estimates to design inductor- and capacitor-based waveform-selective metasurfaces having particular target time constants as well as to predict the transition time between a reflecting state and an absorbing state. However, note that waveform-selective performance possibly becomes poor if the resistance is set to an improper value. For instance, if $R_L$ is too large, then the inductor-based waveform-selective metasurface cannot accept electric charges induced by a long pulse, decreasing its absorptance.

Additionally, our equivalent circuit models enable us to predict entire time-domain response by using eqs. (3) and (6). For instance, FIG. 2(c) and FIG. 2(d) show normalized inductor voltage $\hat{v}_L$ and capacitor voltage $\hat{v}_C$ that were derived analytically and numerically as a function of time. These figures indicate that relatively large difference appeared during an initial time period of the inductor-based waveform-selective metasurface, especially when its inductance increased to a large value (see $L = 1$ mH). This is because the use of large inductance led to enhancing the electromotive force to prevent incoming electric charges, which resulted in more increasing the effective resistance of diodes than assumed in eq. (4) (i.e., more than 680 Ω). Despite such a minor difference, however, these figures demonstrate that the analytical results well characterize the trend of the time-varying responses of both structures.

So far our equivalent circuit models were compared to simulation results with respect to voltages inside a diode bridge. From the practical viewpoint, however, it is more important to evaluate the power of a reflected wave propagating in free space.

This comparison result is seen in FIG. 2(e) and FIG. 2(f), where numerically calculated reflected power is plotted together with a square of voltage obtained from our equivalent circuit models. In these figures, the scale of numerically derived powers is adjusted, as waveform-selective metasurfaces varied their absorptances only between 0.1 and 1.0, while the analytically derived voltages changed between 0.0 and 1.0. As a consequence, these analytical results also closely matched with the numerical results, although relatively large discrepancy appeared for the inductor-based waveform-selective metasurface, since this structure tends to have a large effective diode resistance as explained earlier. We also noticed from FIG. 2(e) and FIG. 2(f) that the reflected power shifted to the right-hand side of each figure, compared to numerically derived squared voltage (FIG. 2(c) and FIG. 2(d)). This is presumably because the electric charges induced by the incident wave did not immediately couple with the lumped circuit components between the conducting patches, which appeared as the delay (shift) in numerically derived reflected power.

The concept of our equivalent circuit approach can be further extended to predicting time-domain responses of more complicated structures such as parallel- and series-type waveform-selective metasurfaces (see the left of FIG. 3(a) and FIG. 3(b)). As explained above, these structures, respectively, reflect and absorb an intermediate pulse more strongly than a short pulse and a CW, which means that their voltages applied inside diode bridges exhibit a peak and a dip in the time domain, respectively[24]. Based on our approach, the circuit models applicable to these structures may be represented as drawn in the right of FIG. 3(a) and FIG. 3(b). In these cases, two sets of the following equations are similarly derived for the parallel- and series-type waveform-selective metasurfaces, i.e.,

$$\begin{cases} i(t) = i_R(t) + i_L(t) + i_C(t) \\ R_d i(t) + R_C i_R(t) = E_0 \\ R_C i_R(t) = R_L i_L(t) + L di_L(t)/dt = C^{-1}\int i_C(t)dt \\ v_P(t) = R_C i_R(t), \end{cases} \quad (9)$$

$$\begin{cases} (R_d + R_L)i(t) + R_C i_R(t) + L di(t)/dt = E_0 \\ R_C i_R(t) = C^{-1}\int i_C(t)dt \\ i(t) = i_R(t) + i_C(t) \\ dq(t)/dt = i_C(t) \\ v_S(t) = R_L i(t) + R_C i_R(t) + L di(t)/dt. \end{cases} \quad (10)$$

In these equations, $v_P$ and $v_S$ are the voltages applied to the entire circuits except diodes. Solving and manipulating the first set of the equations gives us $v_P$ as

$$v_P(t) = \begin{cases} R_C E_0 (R_L R_d + R_L R_C + R_d R_C)^{-1} [R_L \{1 + (\beta_P e^{\alpha_P t} - \alpha_P e^{\beta_P t})/(\alpha_P - \beta_P)\} + L\alpha_P \beta_P (e^{\alpha_P t} - e^{\beta_P t})/(\alpha_P - \beta_P)], \\ \quad \text{for } [\{L(R_d + R_C) + CR_L R_d R_C\}/(LCR_d R_C)]^2 - 4(R_L R_d + R_L R_C + R_d R_C)/(LCR_d R_C) > 0 \\ R_C E_0 (R_L R_d + R_L R_C + R_d R_C)^{-1} [R_L \{1 + e^{h_P t} h_P k_P^{-1} (\sin k_P t - \cos k_P t)\} + L(h_P^2 + k_P^2) k_P^{-1} e^{h_P t} \sin k_P t], \\ \quad \text{for } [\{L(R_d + R_C) + CR_L R_d R_C\}/(LCR_d R_C)]^2 - 4(R_L R_d + R_L R_C + R_d R_C)/(LCR_d R_C) < 0. \end{cases} \quad (11)$$

In this equation, $\alpha_P$ and $h_P$ are the real roots obtained when solving eq. (9) with respect to $i$, while $\beta_P$ and $k_P$ are the imaginary roots. The second set of the above equations can be solved in a similar manner to yield $v_S$, i.e.,

$$v_S(t) = \begin{cases} E_0 (R_d + R_L + R_C)^{-1} [R_C + R_L + (\alpha_S - \beta_S)^{-1} \{(R_C + R_L)(\beta_S e^{\alpha_S t} - \alpha_S e^{\beta_S t}) \\ \quad + \alpha_S \beta_S \langle (L + CR_C R_L)(e^{\alpha_S t} - e^{\beta_S t}) + LCR_C (\alpha_S e^{\alpha_S t} - \beta_S e^{\beta_S t})\rangle\}], \\ \quad \text{for } 4(R_d + R_L + R_C)/(LCR_C) - [\{L + CR_C(R_d + R_L)\}/(LCR_C)]^2 > 0 \\ E_0 (R_d + R_L + R_C)^{-1} [(R_C + R_L) + \{1 - e^{h_S t}(\cos k_S t - h_S k_S^{-1} \sin k_S t)\} \\ \quad + (h_S^2 + k_S^2) k_S^{-1} e^{h_S t} \{\langle CR_C R_L + L(1 + CR_C h_S)\rangle \sin k_S t + LCR_C k_S \cos k_S t\}], \\ \quad \text{for } 4(R_d + R_L + R_C)/(LCR_C) - [\{L + CR_C(R_d + R_L)\}/(LCR_C)]^2 < 0. \end{cases} \quad (12)$$

In this equation, $\alpha_S$ and $h_S$ represent the real roots obtained when solving eq. (10) with respect to $q$, while $\beta_S$ and $k_S$ are the imaginary roots. Equations (11) and (12) indicate that $v_P$ and $v_S$ have different forms depending on the relationship between circuit constants used.

Using equations (11) and (12), time-domain responses of parallel- and series-type waveform-selective metasurfaces were obtained as seen in FIG. 3(c) and FIG. 3(d), where as default values $R_L$, $R_C$, $L$, and $C$ were set to 10 Ω, 10 kΩ, 100 μH, and 100 pF, respectively. These figures show that the analytically derived voltages qualitatively agreed with numerical simulation results, particularly with respect to the locations of voltage peaks and dips, while a relatively large difference was seen for the series-type waveform-selective metasurface, which can be also explained by the influence of effective diode resistance during an initial time period. Similar trend was confirmed, when we alternatively swept the capacitance of the parallel-type waveform-selective metasurface and the inductance of the series-type waveform-selective metasurface (results not shown here).

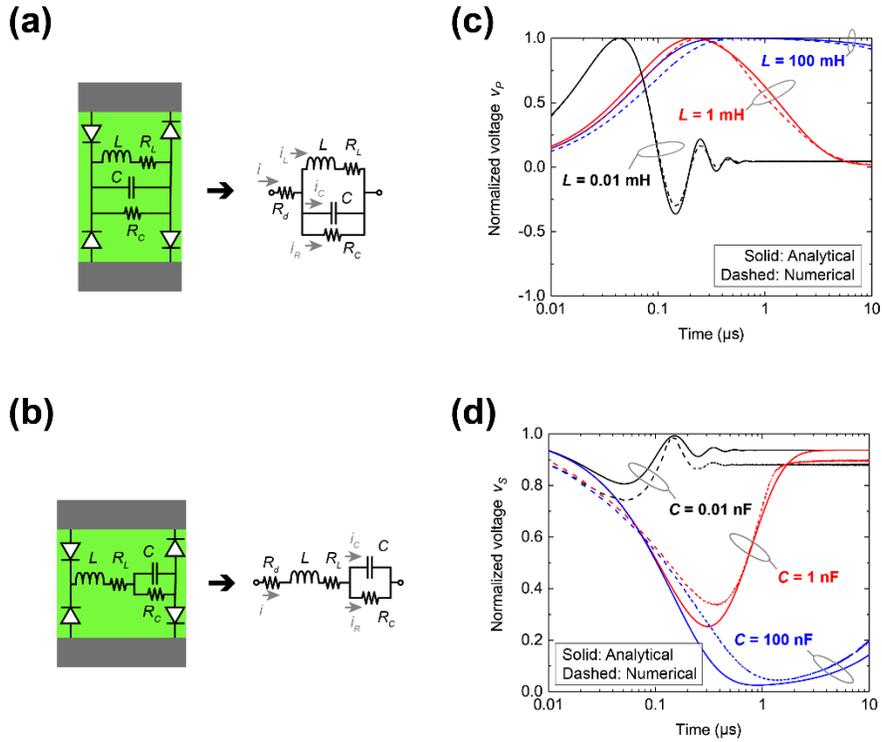

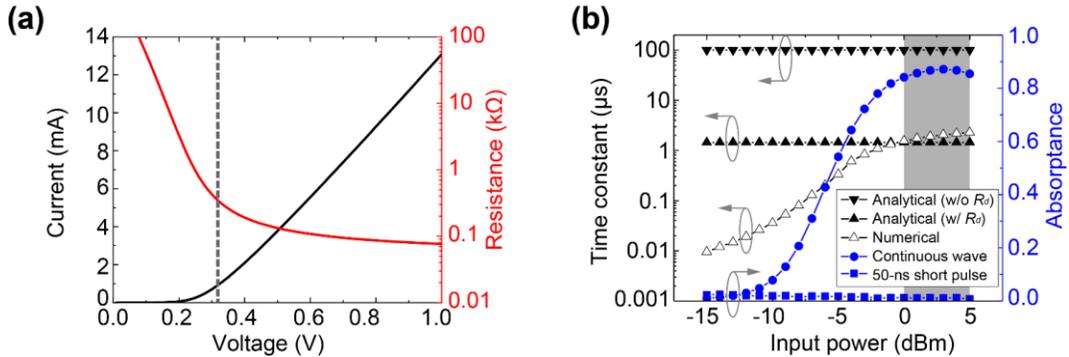

FIG. 3. (a,b) Circuit configurations used for a parallel-type and series-type waveform-selective metasurfaces and proposed equivalent circuit models. (c,d) Their time-domain voltages compared to simulation results.

FIG. 4. (a) Relationship between the voltage and current of the diode used. The red solid curve and gray vertical line represent the effective diode resistance and the turn-on voltage, respectively. (b) Power dependence of the time constant of the inductor-based waveform-selective metasurface used in FIG. 2(a). The blue curves show the corresponding CW and short-pulse absorptances.

As mentioned in FIG. 1(f), our equivalent circuit approach simply set effective diode resistance $R_0$ to be the resistance at turn-on voltage (as marked by the dashed line in FIG. 4(a) and specifically 340 Ω). However, the use of this fixed $R_0$ (or fixed $R_d$) limits predicting the power dependence of the diode (see $R_0$ of FIG. 4(a)) as well as that of waveform selectivity. This influence was evaluated in FIG. 4(b), where the inductor-based waveform-selective metasurface shown in FIG. 2(a) was used again with a fixed inductance of 1 mH but with various input powers. Although the difference between numerical results and analytical results was significantly reduced by introducing effective diode resistance, a relatively large gap appeared by decreasing the input power level, which effectively increased $R_d$ and thus decreased $\tau_L$ in the numerical simulation (see the open triangles in FIG. 4(b)). At the same time, however, this reduced waveform-selective performance as well, namely, the difference between the absorptance for a short pulse and that for a CW (the blue curves of FIG. 4(b)) since the voltage across the diode bridge was not sufficiently large. Importantly, these results imply that still our approach works very well in the practical input power range where waveform-selective absorptance clearly appears (the gray area of FIG. 4(b)). This is because as seen in the right side of the dashed vertical line of FIG. 4(a), the effective diode resistance $R_0$ (and $R_d$) did not significantly change once the diodes were turned on, compared to the left side of the dashed vertical line (also refer to eq. (4) again for the relationship between $R_d$ and $\tau_L$).

To clarify the difference between our metasurfaces and conventional structures, we note again that ordinary metasurfaces vary their behaviors in response to the frequency spectrum of an incoming wave but not to its pulse width at the same frequency. This pulse width dependence is made possible by using our waveform-selective metasurfaces. Such structures give us an additional degree of freedom to control electromagnetic phenomena. Importantly, the conventional design process for waveform-selective metasurfaces was extremely time-consuming since numerical simulations were performed in the time domain until time scale far longer than a period of an incoming wave. Our equivalent circuit approach can be used as first-order design of waveform-selective metasurfaces to reduce the number of numerical simulations and facilitate the design process. Additionally, to more understand waveform-selective mechanisms, for example, eqs. (4), (7), and (8) show when waveform-selective metasurfaces approach their steady states depending on time constants as well as how these values are determined by circuit constants used. Finally, our approach can be more effectively exploited when several types of waveform-selective metasurfaces are nonuniformly integrated. Including such an example, we demonstrated a simplified approach to predict time-domain responses of complicated structures from FIGS. S1 to S3 in Supplementary Material.

In conclusion, we have presented a simplified equivalent circuit approach to estimate time-domain responses of waveform-selective metasurfaces under two assumptions, specifically, simplified input source and diode resistance. Comparison result between analytically and numerically obtained time constants showed that effective diode resistance plays an important role to estimate how a waveform-selective metasurface behaves in the time domain. The concept of our equivalent circuit approach was applied to more complicated types of structures that preferentially absorb or reflect an intermediate pulse at the same frequency, which appears as a voltage dip/peak. As a result, our models showed close agreement with numerical simulation results. Hence, our method is simple yet powerful enough to predict such unusual waveform-selective mechanisms that vary electromagnetic characteristics in the time domain (particularly, in time scale far longer than a period of an incident wave) and to facilitate the design process of the structures.

## SUPPLEMENTARY MATERIAL

See supplementary material for other more complicated examples using our equivalent circuit approach as well as for discussion on effective diode resistance.

## ACKNOWLEDGMENTS

The authors acknowledge K. Ikejiri for valuable discussions and suggestions about turn-on voltage of diodes. This work was supported in part by the Japanese Ministry of Internal Affairs and Communications (MIC) under Strategic Information and Communications R&D Promotion Programme (SCOPE) #165106001 and 192106007 and Japan Science and Technology Agency (JST) under Precursory Research for Embryonic Science and Technology (PRESTO) #JPMJPR193A.

________________________________

[a] K. Asano and T. Nakasha contributed equally to this work.

[b] Electronic mail: wakatsuchi.hiroki@nitech.ac.jp

# Supplementary Material of

*Simplified equivalent circuit approach for designing time-domain responses of waveform-selective metasurfaces*

Kosei Asano,[1,a)] Tomoyuki Nakasha,[1,a)] and Hiroki Wakatsuchi[1,2,b)]

[1]*Department of Electrical and Mechanical Engineering, Graduate School of Engineering, Nagoya Institute of Technology, Aichi, 466-8555, Japan*

[2]*Precursory Research for Embryonic Science and Technology (PRESTO), Japan Science and Technology Agency (JST), Saitama 332-0012, Japan*

[a]*These authors contributed equally to this work.*

[b]*Electronic mail: wakatsuchi.hiroki@nitech.ac.jp*

In FIG. 3 we showed entire time-domain responses of parallel- and series-type waveform-selective metasurfaces using eqs. (11) and (12). By differentiating these equations and equating them to zero, we can predict their reflectance peak or dip in the time domain, which is one of important aspects of these structures to selectively control intermediate pulses, although this approach may be a little complicated.

Alternatively, the time-domain responses of these structures may be more easily estimated by using separated individual inductor- and capacitor-based waveform-selective metasurfaces that have the same circuit values as the ones incorporated in the combined structures. Basically, parallel- and series-type waveform-selective metasurfaces behave like a combination of these individual structures. In this case, a voltage peak/dip may be briefly estimated by the intersection of voltage curves calculated by eqs. (3) and (6). This is demonstrated in FIG. S1, where normalized $v_P$ of a parallel-type waveform-selective metasurface and normalized $v_S$ of a series-type waveform-selective metasurface were analytically and numerically obtained. In addition, this figure shows normalized $v_L$ of an inductor-based waveform-selective metasurface and normalized $v_C$ of a capacitor-based waveform-selective metasurface. Note that these structures used the same circuit constants, specifically, $L = 100$ μH, $C = 1$ nF, $R_L = 10$ Ω, and $R_C = 10$ kΩ. Under these circumstances, the intersection of the analytically calculated $v_L$ and $v_C$ appeared around 0.2 μs, which was close to the intersection of the numerically obtained values. Moreover, they were found to be near the first voltage peak of the parallel-type waveform-selective metasurface and the voltage dip of the series-type waveform-selective metasurface (both between 0.3 and 0.4 μs).

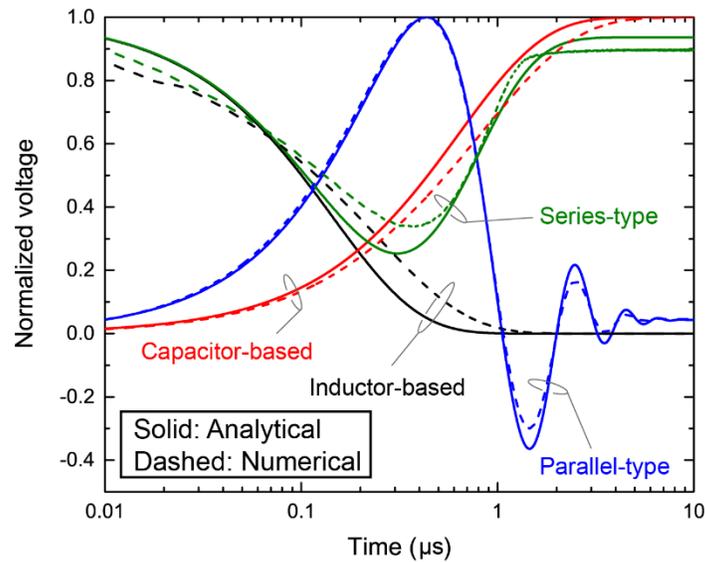

FIG. S1. Normalized voltages of inductor- and capacitor-based and parallel- and series-type waveform-selective metasurfaces using the same circuit constants ($L = 100$ μH, $C = 1$ nF, $R_C = 10$ kΩ, and $R_L = 10$ Ω).

This approach may be applicable to more complicated cases as well. For instance, FIG. S2(a) shows a structure combining the circuit configuration of a parallel-type waveform-selective metasurface ($L = 1$ mH, $C = 0.1$ nF, $R_L = 10$ Ω, and $R_C = 10$ kΩ) with that of a capacitor-based waveform-selective metasurface ($C_2 = 10$ nF, $R_C = 10$ kΩ) in series[36]. As plotted in FIG. S2(b), such a structure exhibited a voltage peak like the ones seen in the blue curves of FIG. S1 (i.e., a parallel-type waveform-selective metasurface) but then started increasing the voltage again. Additionally, this figure plots the voltages of decomposed individual structures (i.e., two capacitor-based waveform-selective metasurfaces, each using $C/C_2 = 0.1/10$ nF and $R_C = 10$ kΩ, and one inductor-based waveform-selective metasurface using $L = 1$ mH and $R_L = 10$ Ω). According to these results, the first intersection of these individual structures (the black and red curves in FIG. S2(b)) was obtained between 0.1 and 0.2 μs, which was close to the first peak of the combined structure (between 0.2 and 0.3 μs). Additionally, the second intersection of these individual structures (the red and gray curves) appeared around 2 µs, which was also near the dip of the combined structure (between 3 and 4 µs).

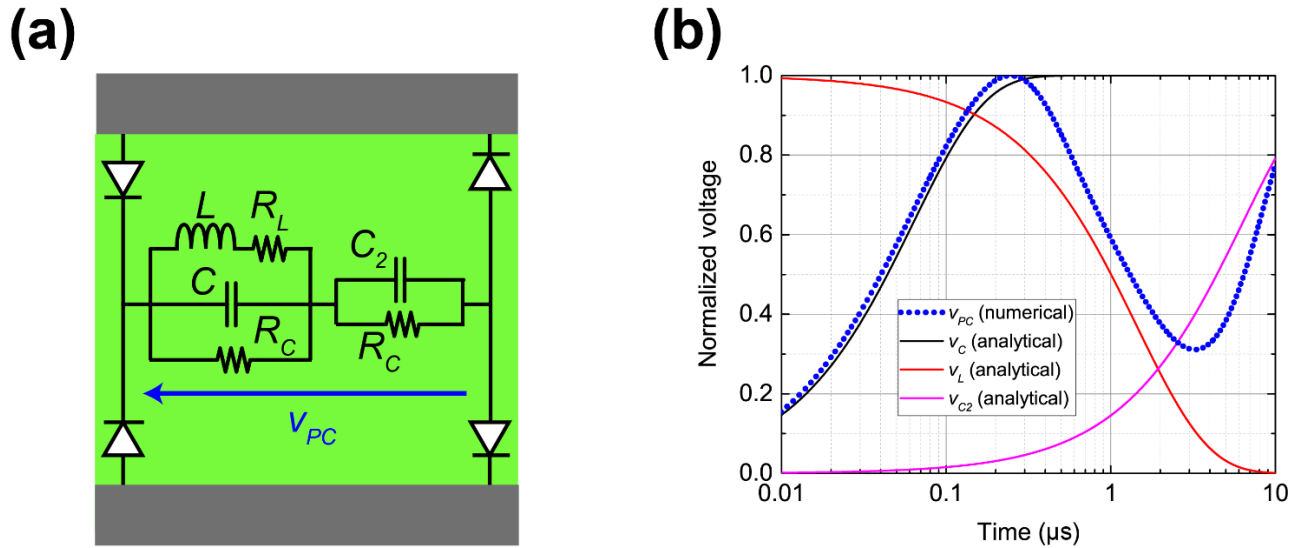

FIG. S2. (a) A structure integrating the circuit configuration of a parallel-type waveform-selective metasurface with that of a capacitor-based waveform-selective metasurface in series. (b) Its normalized voltage compared to those of individual waveform-selective metasurfaces.

Another method to analytically estimate the voltage peaks and dips of FIGS. S1 and S2 is to equate the right side of eq. (3) to the right side of eq. (6), which yields the above intersections as well. However, this approach may escalate a complexity in solving the equation at the same time.

FIG. S3(a) introduces another complicated case where unit cells of an inductor-based waveform-selective metasurface and those of a capacitor-based waveform-selective metasurface were alternately and periodically deployed ($L = 1$ mH, $C = 0.1$ nF,

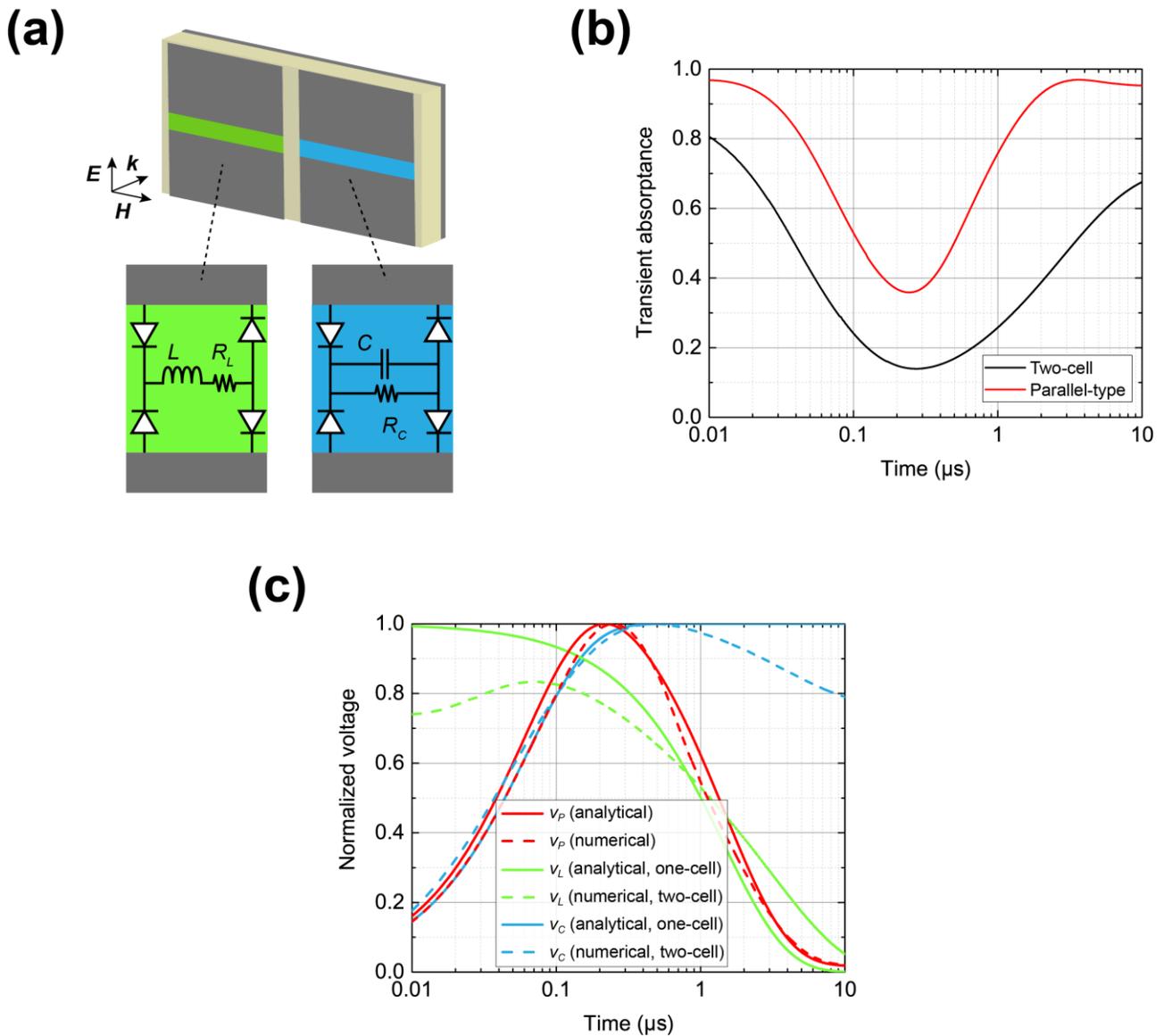

FIG. S3. (a) A two-cell model comprising a unit cell of an inductor-based waveform-selective metasurface and that of a capacitor-based waveform-selective metasurface. (b) Transient absorptances of the two-cell model and a parallel-type waveform-selective metasurface using the same circuit constants. (b) Their normalized voltages compared to those of individual inductor- and capacitor-based waveform-selective metasurfaces (indicated as "analytical, one-cell").

S3

$R_L = 10\ \Omega$, and $R_C = 10\ k\Omega$). Basically, such a structure behaves like a parallel-type waveform-selective metasurface as both a short pulse and a CW are effectively absorbed. However, entirely the absorbing performance is reduced because of shrunk effective absorbing area of each unit cell, which is demonstrated in FIG. S3(b). Also, compared to $v_P$ of a parallel-type waveform-selective metasurface, which exhibits a peak value in the time domain, this structure showed gradually decreased $v_L$ and increased $v_C$, although these voltages slightly reduced during an initial time period and at a steady state, respectively, due to the interaction between different unit cells, as plotted in FIG. S3(c). FIGS. S3(b) and S3(c) show that the minimum transient absorptance of this complicated structure appeared in proximity to the intersection of the numerically obtained $v_L$ and $v_C$ as well as to that of the analytically derived $v_L$ and $v_C$ of separated individual structures. These results support that the smallest absorptance can be almost predicted by using our proposed approach.

Finally, $R_0$ (and $R_d$) depends on what diodes are specifically used (in our case, HSMS286x series of Broadcom). This resistance value may be obtained numerically or experimentally. For example, a current-voltage curve (like the one seen in FIG. 4(a)) can be calculated from a SPICE model or an actual product (or from its datasheet). Approximating the curve by a straight line starting from a sufficiently large voltage gives a good estimate of a turn-voltage.